\documentclass[aps,prl,twocolumn,groupedaddress,noshowpacs,floatfix]{revtex4-1}

\usepackage{graphicx}
\usepackage{dcolumn}
\usepackage{bm}
\usepackage{amsmath}

\begin{document}

\noindent
{\large{\bf Comment on ``Observation of a Pinning Mode in a Wigner Solid with $\nu=1/3$
Fractional Quantum Hall Excitations''}}\\ 

The appearance of disorder-pinned Wigner crystalline phases flanking the low-$\nu$ 
fractional quantum Hall effect (FQHE) states \cite{li00,ye02,chen04} (seen for 
$\nu \leq 0.218$ \cite{chen04}) is well known. These observations have been attributed 
to the presence of strong crystalline correlations in the FQHE states \cite{yl04,jain06}. 
Exact diagonalization (EXD) calculations have shown \cite{yl04} that such crystalline 
correlations maintain for even larger fractions, including $\nu = 1/3$. A recent letter 
\cite{zhu10} reports the first measurements of a pinned mode in a Wigner solid at the
(electronic) $\nu = 1/3$ FQHE filling. The authors \cite{zhu10} suggest that an 
interpretation of their findings may be found within interacting composite fermion (ICF) 
theory. However, to date there is no record reporting crystalline correlations for  
$\nu = 1/3$ obtained from ICF theory \cite{jain06,jainbook} (see also Refs.
23 and 24 in Ref.\ \cite{zhu10}). 

\begin{figure}[t]
\centering\includegraphics[width=4.5cm]{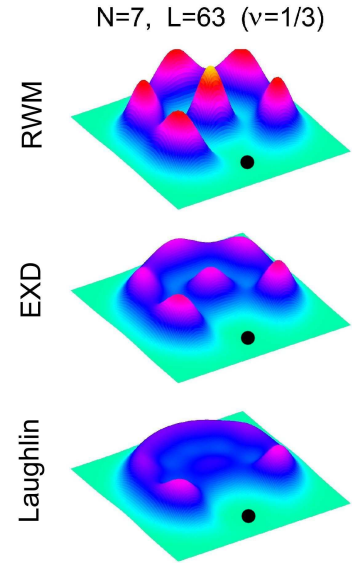}
\caption{
CPDs at fractional filling $\nu=1/3$ for $N=7$ electrons [corresponding to a total angular 
momentum $L=63$; $\nu=N(N-1)/2L$] in the lowest Landau level. Three levels of theoretical 
description are presented: variational RWM, EXD, and variational Laughlin wave function 
(mean-field composite fermion); for the meaning of the labels, see the text. The solid dots 
mark the position of the fixed point ${\bf r}_0$.
} 
\label{cpds}
\end{figure}

In Fig. \ref{cpds} we show for $\nu = 1/3$ the conditional probability distributions 
(CPDs) for three theoretical descriptions: (a) the variational rotating Wigner molecule 
(RWM) approach, (b) an EXD calculation in the lowest Landau level, and (c) a calculation 
using the variational Laughlin wave function; the CPD, 
$P({\bf r}, {\bf r}_0)$, gives the probability 
for finding a particle at position {\bf r} given that another one is located at ${\bf r}_0$ 
\cite{yl04}. The appearance of Wigner crystalline correlations in the EXD is apparent. The 
liquid Laughlin description (coinciding with the mean-field CF approach) fails to reproduce
the EXD crystalline correlations. The variational RWM, while reproducing qualitatively the 
crystalline features of the EXD results, overestimates somewhat the localization of the 
electrons relative to each other. Recently it has been demonstrated that quantitative 
agreement (to machine precision) between the results of EXD calculations and those 
obtained by the RWM method can be obtained by inclusion of the ro-vibrational spectrum 
of the RWM  \cite{yl10}, thus providing a unified physical and mathematical picture which 
encompasses both the crystalline and liquid aspects in the FQHE regime (in particular for 
$\nu=1/3$, as well as in its neighborhood created by deviations of $\nu$ from 1/3 \cite{note}, 
as revealed experimentally \cite{zhu10}).

This research was supported by the US D.O.E. (Grant No. FG05-86ER45234).\\
~~~~~~~~~\\
~~~~~~~~~\\
Constantine Yannouleas and Uzi Landman\\
School of Physics\\ 
Georgia Institute of Technology\\ 
Atlanta, GA 30332-0430, USA\\
20 September 2010


\begin{thebibliography}{99}
\bibitem{note}
The rovibrational RWM excitations carry angular momentum $\lambda$, and thus the RWM excited
states have total angular momentum $L=L_{\text{mag}}+\lambda m$, where $L_{\text{mag}}$ 
denotes magic angular momenta associated with the RWM ground states and $m$ is the 
multiplicity of the vibrational state. The difference $L-L_{\text{mag}}$ corresponds to 
deviations of $\nu$ from the magic fractions such as 1/3. The existence of crystalline 
correlations in the ground state directly implies the presence of crystalline correlations also
in the excited states as demonstrated explicitly in Ref. \cite{yl10}.
\end{thebibliography}
\end{document}